\documentclass[prd,twocolumn]{revtex4}
\usepackage{dcolumn}
\usepackage{multirow}
\usepackage{graphicx}
\usepackage{amssymb}
\usepackage{bm}
\usepackage{hyperref}
\usepackage{epstopdf}
\usepackage{color}
\usepackage{mathrsfs}
\usepackage{amsmath,amssymb,amsthm}
\begin{document}
\title{Cosmological forecasts from current observations of LIGO}

\author{Deng Wang}

\email{Cstar@mail.nankai.edu.cn}
\affiliation{Theoretical Physics Division, Chern Institute of Mathematics, Nankai University,
Tianjin 300071, China}
\author{Xin-He Meng}
\affiliation{{Department of Physics, Nankai University, Tianjin 300071, China}}
\begin{abstract}
We use the simulated gravitational-wave data to explore the evolution of the universe in light of current observations of the Laser Interferometer Gravitational-Wave Observatory (LIGO).  Taking advantage of state-of-the-art Markov Chain Monte Carlo technique to constrain the basic cosmological parameters, the Hubble constant, present matter density parameter and equation of state of dark energy, we find that LIGO needs about, at least 5-year data accumulation, namely about 1000 events, to achieve the accuracy comparable to the Planck result. We also find that, from a new information channel, the constrained value of the Hubble constant from 1000 simulated events is more consistent with the direct local observation by Riess {\it et al.} than the indirect global measurement by the Planck Collaboration at the $2\sigma$ confidence level. The combination of gravitational waves and electromagnetic signals is very prospective to reveal the underlying physics of the universe.
\end{abstract}
\maketitle
\section{Introduction}
In the past almost two decades, the late-time acceleration of the universe has been discovered and confirmed by a large number of cosmic probes such as Type Ia supernovae (SNIa) \cite{1,2}, cosmic microwave background (CMB) radiation \cite{3,4}, large scale structure (LSS) observations \cite{5}, weak gravitational lensing \cite{6} and so on. To explain this accelerated mechanism, physicists introduce phenomenologically a cosmic fluid with abnormal negative pressure dubbed dark energy (DE). Up to date, the most popular cosmological scenario to explain the intriguing phenomena is still the simplest one, i.e., a combination of the cosmological constant $\Lambda$ and cold dark matter (CDM) component ($\Lambda$CDM model). Especially, the Planck-2015 public release with an unprecedented accuracy verified, once again, the correctness of the standard six-parameter $\Lambda$CDM cosmology \cite{7}. To a large extent, this result enhances our confidence to the current standard cosmological paradigm. Nonetheless, this model is not as perfect as one can expect and encounters several severe challenges: (i) the so-called fine-tuning and coincidence problems \cite{8}; (ii) the small scale puzzles of CDM \cite{9}; (iii) the Hubble constant tension over $3\sigma$ confidence level (CL) between the direct local observation from Riess {\it et al.} \cite{10} using improved SNIa calibration techniques and indirect global measurement from the Planck Collaboration under the assumption of $\Lambda$CDM \cite{11}; (iv) the inconsistencies of the amplitude of matter density fluctuations between the Planck CMB data and some low redshift surveys including  lensing, cluster counts and redshift space distortions (RSD) \cite{11,12,13}. These discrepancies indicate that the $\Lambda$CDM model still needs to be further validated at small and large scales. Meanwhile, determining the basic cosmological parameters such as the Hubble constant and present matter density with high accuracy is one of the most urgent tasks in modern observational cosmology. Furthermore, one may naturally ask whether the late universe is dominated by the dynamical dark energy (DDE) at all, whose equation of state (EoS) is $\omega\neq-1$. Therefore, it is also important to determine the value of EoS of DE with forthcoming high-precision data.

Until now, one should notice that main observations (SNIa, CMB and LSS) to explore the background evolution of the universe are based on the electromagnetic (EM) measurements. However, the observation of gravitational wave (GW) will open a new information channel to constrain the cosmological quantities. Based on the fact that the physics governing the inspiral of a binary due to GW emission is well characterized in the framework of general relativity, Schutz \cite{14} first proposed that it is possible to implement a direct and absolute measurement of the luminosity distance to a source by measuring the gravitational waveform during the inspiral and merger of a binary. Standard sirens are GW sources for which the redshift and luminosity distance can be determined, and consequently are the GW analog to standard candles \cite{15,16,17,18,19,20,21,22,23}. Making use of the simplicity of black holes (BH), which are completely described by mass, spin and charge, gravitational sirens can give luminosity distances without invoking the cosmic distance ladders or phenomenological scaling relations. Note that the GW sources do not provide the redshift to the source, since the redshift is highly degenerate with the intrinsic parameters of the sources. Determining accurately the redshift related to a GW source is quite a challenge in the field of GW astronomy. In the literature, there are several scenarios to address this topic such as the use of neuron star mass distribution \cite{24}, the utilization of the tidal deformation of neutron stars \cite{25} and the identification of an accompanying EM signal \cite{19,20,21,22}. Once the redshift of a source is determined in a statistical fashion, we can fully unlock the power of the binary sources as cosmological standard sirens in revealing the evolution of the universe.

A century after Einstein¡¯s prediction about the existence of the GW, on September 14, 2015 at 09:50:45 UTC the two detectors of the LIGO simultaneously observed the first transient GW signal GW150914 in human history \cite{26}. This opens the door to new areas of study not possible with light given the electromagnetically quiet nature of BH mergers, and marks the beginning of a new era of multi-messenger astronomy. What is more exciting, the LIGO collaboration reported subsequently two GW events GW151226 and GW170104 from the coalescence of two stellar-mass binary black holes (BBH) \cite{27,28}, which were identified with high statistical significance, as well as a candidate LVT151012 also probable to be a BBH system \cite{29,30}. Based on these observations, the LIGO collaboration predicts that the merger rate of BBH system will lie in the range 9-240 Gpc$^{-3}$yr$^{-1}$ \cite{30,31}. If assuming tentatively LVT151012 is also a realistic GW event, we can find that the LIGO detected four events in about a year and a half. This implies that with increasing sensitivity, the LIGO have enough potential to detect more GW events an provide more abundant information for us. In light of current detection sensitivity of LIGO, the aim of this work is to make forecasted constraints on the basic cosmological parameters under the most optimistic assumption by using the current sparse GW data. We find that LIGO needs about at least 5-year data accumulation to achieve the accuracy comparable to the Planck results.

The structure of this work is displayed as follows. In the next section, we introduce the basic cosmological formula. In Section III, we present our simulation based on the current four GW events, while we implement the numerical analysis and exhibit our results in Section IV. The discussions and conclusions are presented in the final section.

\section{Background formula}
In a Friedmann-Robertson-Walker (FRW) universe, the luminosity distance $d_L(z)$ can be written as
\begin{equation}
d_L(z)=\frac{c(1+z)}{H_0\sqrt{|\Omega_{k0}|}}sinn\left(\sqrt{|\Omega_{k0}|}\int^{z}_{0}\frac{dz'}{E(z')}\right), \label{1}
\end{equation}
where $z$ denotes the redshift, $c$ is the speed of light, the dimensionless Hubble parameter $E(z)=H(z)/H_0$, the present cosmic curvature $\Omega_{k0}=-Kc^2/(a_0H_0^2)$, and for $sinn(x)= sin(x)$, $x$, $sinh(x)$, $K=1$, 0, $-1$, which corresponds to a closed, flat and open universe, respectively. Since we just focus on the evolution of the late universe and the Planck CMB data has given a very stringent constraint on the present cosmic curvature $\Omega_{k0}<|0.005|$ \cite{7}, we ignore the background contribution from the relativistic radiation and curvature components in the cosmic pie. To estimate the abilities of current LIGO experiment in describing the evolution of the universe, we use the simulated GW data to constrain three cosmological models: $\Lambda$CDM, $\omega$CDM, and Chevallier-Polarski-Linder (CPL) parametrization \cite{32,33}. The squared dimensionless Hubble parameter of the $\Lambda$CDM model reads
\begin{equation}
E^2(z)=\Omega_{m0}(1+z)^3+1-\Omega_{m0}, \label{2}
\end{equation}
where $\Omega_{m0}$ is the present matter density parameter. The $\omega$CDM model regarding the DE as a single negative pressure fluid is the simplest parametrization of EoS of DE $\omega(z)=\omega=constant$, and its squared dimensionless Hubble parameter of $\omega$CDM model is written as
\begin{equation}
E^2(z)=\Omega_{m0}(1+z)^3+(1-\Omega_{m0})(1+z)^{3(1+\omega)}. \label{3}
\end{equation}
If $\omega$ differs from -1, it is very likely to evolve with time. The CPL model $\omega(z)=\omega+\omega_a\frac{z}{1+z}$ gives an excellent fit for a great deal of theoretically conceivable scalar field potential scenarios and provide a  a good explanation for small deviation from the phantom divide ($\omega=-1$). Meanwhile, $\omega(z)$ is also a well-behaved function at $z\gg1$ and recovers the linear behavior at low redshifts. The squared dimensionless Hubble parameter of the CPL model can be expressed as
\begin{equation}
E^2(z)=\Omega_{m0}(1+z)^3+(1-\Omega_{m0})(1+z)^{3(1+\omega+\omega_a)} e^{-\frac{3\omega_a z}{1+z}}. \label{4}
\end{equation}
where $\omega$ and $\omega_a$ are two free parameters of CPL model. One can easily find that when $\omega=-1$ in Eq. (\ref{3}) and $\omega=-1$ and $\omega_a=0$ in Eq. (\ref{4}), the $\omega$CDM and CPL models reduce to the $\Lambda$CDM one (see Eq. (\ref{2})), respectively.
\begin{table}[h!]
\renewcommand\arraystretch{1.3}
\begin{tabular}{ccccccc}
\hline
\hline
      Event                     &Redshift $z$ \quad  &Luminosity distance $d_L/$Mpc  \\
\hline
GW150914             &$0.09^{+0.03}_{-0.04}$        &$420^{+150}_{-180}$                         \\
LVT151012                    &$0.20^{+0.09}_{-0.09}$       &$1000^{+500}_{-500}$                      \\
GW151226                   &$0.09^{+0.03}_{-0.04}$       &$440^{+180}_{-190}$                       \\
GW170104                     &$0.18^{+0.08}_{-0.07}$       &$880^{+450}_{-390}$                      \\
\hline
\hline
\end{tabular}
\caption{The current available 4 GW data points.}
\label{t1}
\end{table}

\section{GW simulation}
\begin{figure}
\centering
\includegraphics[scale=0.5]{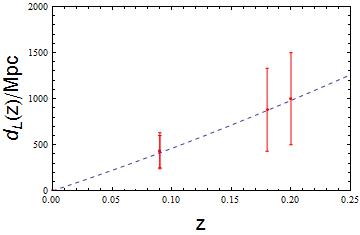}
\caption{The relation between the luminosity distances of basic 5 GW data points with red error bars and redshift $z$ is shown. The blue dashed line corresponds to the $H_0=70$ km s$^{-1}$ Mpc$^{-1}$, $\Omega_{m0}=0.3$ standard cosmology.}\label{f1}
\end{figure}

As emphasized above, we implement tentatively the cosmological forecasts according to current GW observations of LIGO in this section. We list the available 4 GW data points in Tab. \ref{t1}. Note that here we have assumed LVT151012 as a realistic GW event to carry out the simulation. Through observing the data, one can find that the redshifts of 4 GW sources have large uncertainties and these error bars are almost symmetric about the best-fit redshifts. Since current GW data is very sparse and our goal is to estimate the cosmological quantities under the most optimistic assumption, we just consider the best-fit redshift points for every GW event. Although the Fermi Gamma-ray Burst Monitor (GBM) and Large Area Telescope (LAT) observations of LIGO GW events still indicate no evidence of EM counterparts \cite{34,35,36}, we cannot rule out the possibility of detecting accompanying EM signals with gradually improved observational techniques in the future. As a consequence, we may reasonably assume that future experiments can measure the redshifts of 4 GW sources with ultrahigh precision by identifying them as realistic EM counterparts (e.g. short and intense gamma-ray bursts). Furthermore, these measured redshifts with high accuracy can be reasonably regarded as the best-fit redshifts of 4 GW events. In addition, one can also observe that the luminosity distances of 3 GW events obey asymmetric uncertainties. For simplicity, we use the maximal ones of the upper and lower limits of data uncertainties as the errors of luminosity distances. By adding a boundary data point $d_L(0)=0$ into the above GW data, we show our basic 5 GW data points before the simulation in Fig. \ref{f1}.
\begin{figure}
\centering
\includegraphics[scale=0.5]{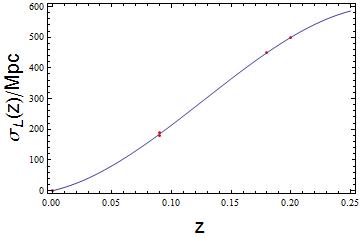}
\caption{The relation between the errors of the luminosity distances of basic 5 GW data points and redshift $z$ is shown. The blue solid line corresponds to the linear cubic polynomial function $\sigma_L(z)=-44893.4z^3+17059.5z^2+883.8z-7.7\times10^{-14}$. }\label{f2}
\end{figure}
\begin{figure}
\centering
\includegraphics[scale=0.5]{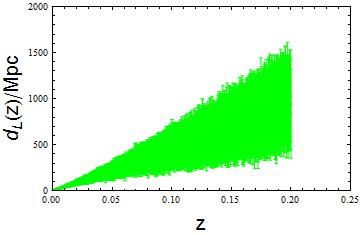}
\caption{The 1000 simulated GW events in light of current observations of LIGO.}\label{f3}
\end{figure}
\begin{figure}
\centering
\includegraphics[scale=0.5]{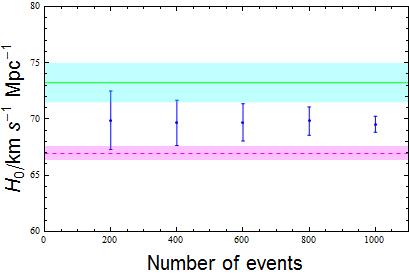}
\caption{In the framework of $\Lambda$CDM model, the relation between the MCMC analysis results of the Hubble constant $H_0$ with $68\%$ CL ranges (blue error bars) and different numbers of GW events is shown. The dashed magenta line and pink band are the best-fit value and $68\%$ CL range of Planck-2015 CMB analysis result. The green solid line and cyan band are the best-fit value and $68\%$ CL range of Riess {\it et al.} result. }\label{f4}
\end{figure}
\begin{figure}
\centering
\includegraphics[scale=0.5]{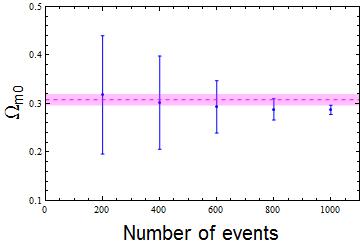}
\includegraphics[scale=0.51]{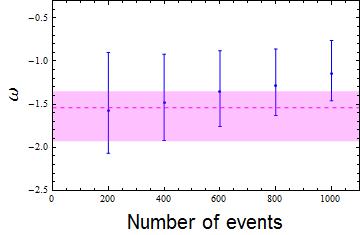}
\caption{\textit{Upper panel}: The relation between the MCMC analysis results of the present matter density parameter $\Omega_{m0}$ with $68\%$ CL ranges and different numbers of GW events is shown. The dashed magenta line and pink band are the best-fit value and $68\%$ CL range of Planck-2015 CMB analysis result. \textit{Lower panel}: The relation between the MCMC analysis results of EoS of DE $\omega$ with $68\%$ CL ranges and different numbers of GW events is shown. The dashed magenta line and pink band are the best-fit value and $68\%$ CL range of Planck-2015 CMB analysis result using Planck TT + lowP. }\label{f5}
\end{figure}
\begin{figure}
\centering
\includegraphics[scale=0.55]{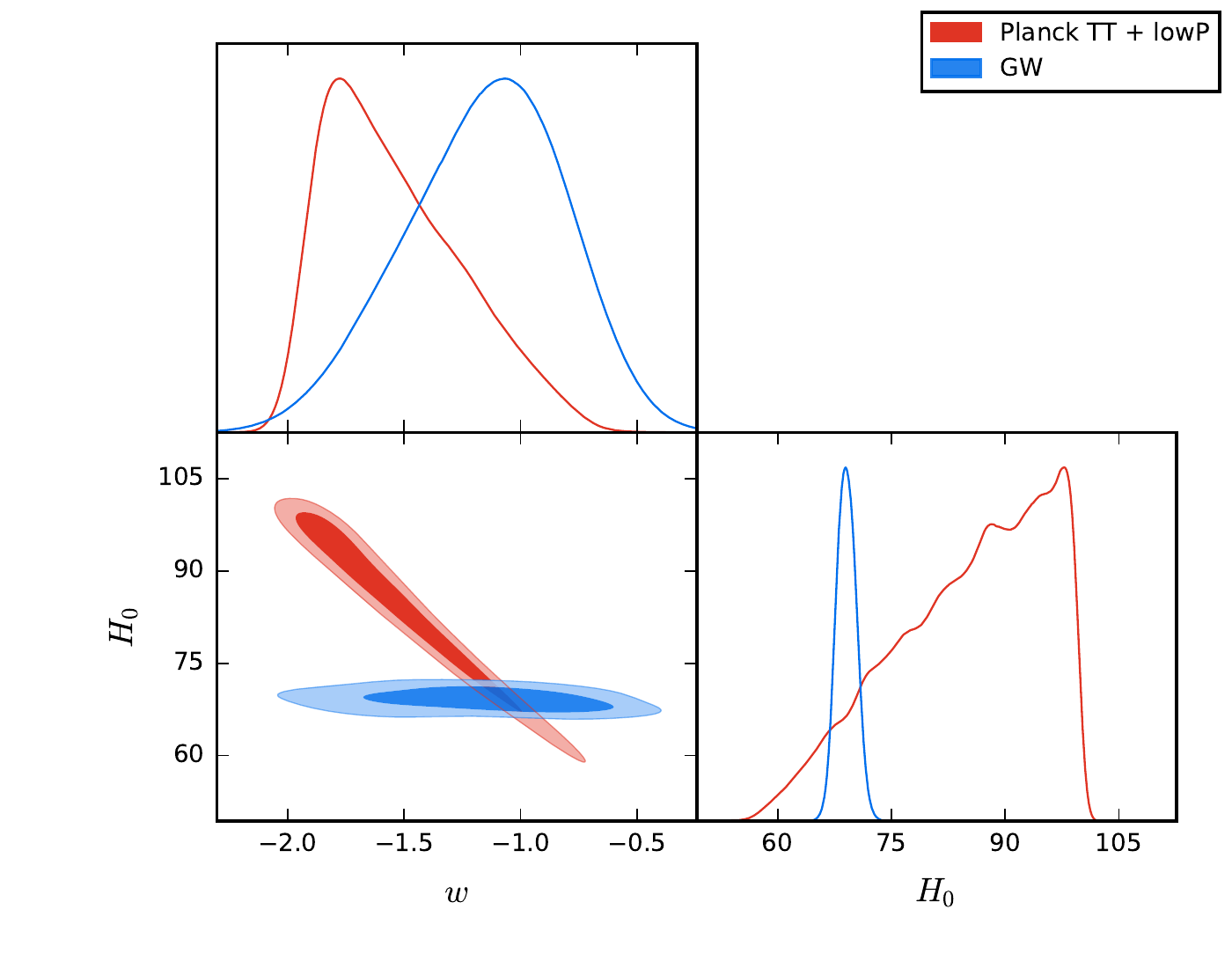}
\caption{The 1-dimensional and 2-dimensional marginalized posterior distributions for $\omega$CDM model using the 1000 simulated GW events and Planck TT + lowP, respectively. }\label{f6}
\end{figure}
\begin{figure}
\centering
\includegraphics[scale=0.6]{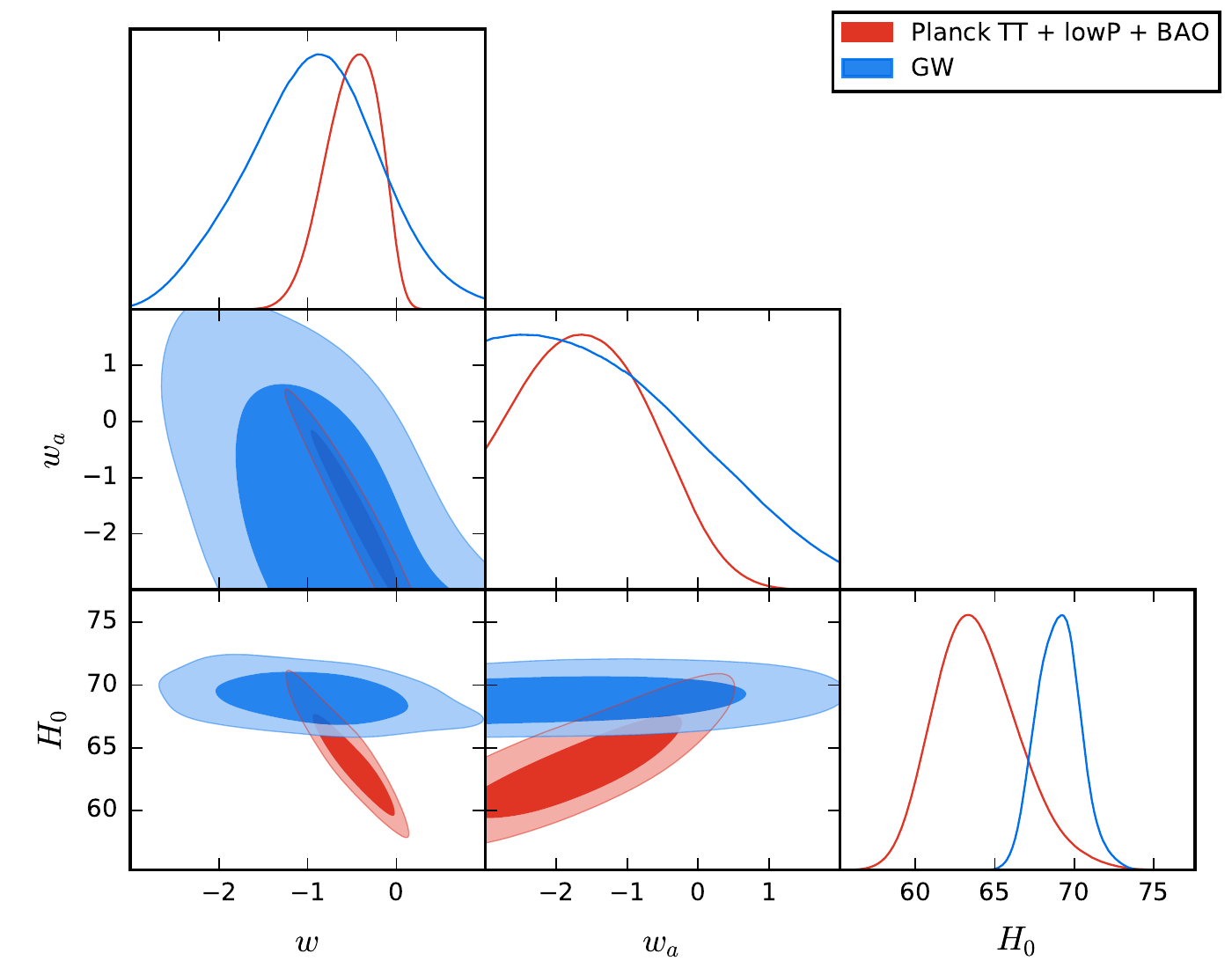}
\caption{The 1-dimensional and 2-dimensional marginalized posterior distributions for CPL model using the 1000 simulated GW events and Planck TT + lowP + BAO, respectively.}\label{f7}
\end{figure}

Due to the lack of redshift distribution information of GW sources and the fact the largest redshift is up to 0.2 (LVT151012), we run Monte Carlo simulations and generate synthetic samples of $d_L(z)$ in the redshift range $z\in [0, 0.2]$, where the redshifts of simulated GW events obey the simplest uniform distribution. To implement a concrete simulation, we choose the $H_0=70$ km s$^{-1}$ Mpc$^{-1}$, $\Omega_{m0}=0.3$ standard cosmology as our fiducial model (see Fig. \ref{f1}).   By calculating the relative difference ratios of the best-fit luminosity distances of 4 GW events with respect to the corresponding values of the fiducial model at the best-fit redshifts (e.g. $(880-d_L(0.18))/d_L(0.18)=0.88\%$ for GW170104), we find that GW151226 gives the largest relative difference ratio (RDR) $6.86\%$. Consequently, we set a upper limit $6.86\%$ for the RDR of luminosity distance in the process of simulation. Then, we fit a linear cubic polynomial function (LCPF) $\sigma_L(z)=-44893.4z^3+17059.5z^2+883.8z-7.7\times10^{-14}$ to the basic 5 GW data points in order to ensure the redshift satisfying $\sigma_L(z)=0$ is smaller than the redshift of the first simulated event (see Fig. \ref{f2}). For instance, if we simulate 200 GW events in the case of LCPF, the redshift $z=8.71\times10^{-17}<0.001$ is well satisfied. However, if we consider the case of a linear square polynomial function, the redshift $z=0.009$ letting $\sigma_L(z)=0$ is larger than 0.001. Therefore, the LCPF is a very good choice for our simulation. Meanwhile, as similarly as done for the RDR of luminosity distance, we find that GW151226 gives the largest RDR of the error of luminosity distance $2.70\%$ (e.g. $(190-d_L(0.09))/d_L(0.09)=2.70\%$) and also set another upper bound on the RDR of the error of luminosity distance $2.70\%$ in our simulation.

The simulated GW samples are generated by using $d_{Lsim}(z)=d_{Lfid}(z)+N(0,\tilde{{\sigma_L}}(z))$, where $d_{Lsim}(z)$, $d_{Lfid}(z)$ and $N(0, \tilde{{\sigma_L}}(z))$ denote the simulated values of luminosity distance at redshift $z$, fiducial values of luminosity distance at redshift $z$ and random numbers normally distributed with mean zero and variance $\tilde{{\sigma_L}}(z)$, respectively. The errors of the simulate data $\tilde{{\sigma_L}}(z)$ also obey a normal distribution $N(0, \sigma_{ero}(z))$. It is worth noting that we have discarded the data points which go beyond the RDR of luminosity distance $6.86\%$ for $\tilde{{\sigma_L}}(z)$ and that of the error of luminosity distance $2.70\%$ for $\sigma_{ero}(z)$. We repeat this process $10^4$ times, and calculate the mean values of the luminosity distance and its related error at each redshift, respectively. Hence, these mean values can be regarded as our simulated GW data.

Since the LIGO collaboration reported the merger rate of GW events lies in the range 9-240 Gpc$^{-3}$yr$^{-1}$, we make the most optimistic assumption in light of current sensitivity of LIGO that the LIGO will observe 200 events per Gpc$^{-3}$ per year. We are aimed at exploring how many future identified GW events can give the cosmological estimations comparable to recent results by the Planck Collaboration. To this end, we simulate 200 (1 year), 400 (2 year), 600 (3 year), 800 (4 year) and 1000 (5 year) events, respectively. As a performance, we show the simulated 5-year data, namely 1000 events in Fig. \ref{f3}. To constrain the basic cosmological quantities, we adopt a usual $\chi^2$ statistics for the simulated GW data as follows
\begin{equation}
\chi^2=\sum^{n}_{i=1}[\frac{\bar{d_L}(z_i)-d_L(z_i;\vec{\theta})}{\bar{\sigma_L}(z_i)}]^2, \label{5}
\end{equation}
where $\bar{d_L}(z_i)$ and $\bar{\sigma_L}(z_i)$ are the luminosity distance and corresponding $1\sigma$ error of luminosity distance for the simulated GW data at a given redshift $z_i$, $\vec{\theta}$ denote free parameters of different cosmological models, and $n$ is number of simulated events, respectively.
\section{Analysis results}
Utilizing the simulated GW data, we employ the Markov Chain Monte Carlo (MCMC) method to constrain the basic cosmological parameters. More specifically, we use the public MCMC package CosmoMC as a sampler \cite{37}, which obeys a convergence diagnostic based on the Gelman and Rubin statistic. To perform the Bayesian
analysis, we choose the uniform priors for different model parameters as follows: $\Omega_{m0} \in [0.01,0.9]$, $\omega \in [-3, 1]$, and $\omega_a \in [-3, 3]$. It is noteworthy that the prior ranges for different model parameters are chosen to be much wider than the posterior ones in order not to affect the results of parameter estimation.

In Fig. \ref{f4}, we show our MCMC estimations of $H_0$ for different numbers of events in the framework of $\Lambda$CDM model and make a comparison with the results of Planck-2015 CMB and Riess {\it et al.} analysis. We find that about 600 events give the same accuracy of $H_0$ as Riess {\it et al.} result \cite{10}, who recently give a $2.4\%$ local determination on $H_0$, and that 1000 events can give $H_0=69.54\pm0.71$ km s$^{-1}$ Mpc$^{-1}$ with $1\%$ precision, which is well comparable to the Planck-2015 result $H_0=66.93\pm0.62$ km s$^{-1}$ Mpc$^{-1}$ with $0.9\%$ precision \cite{11}. Interestingly, we also find that these 1000 simulated GW data points can alleviate the current $H_0$ tension from 3.4$\sigma$ to 1.97$\sigma$ CL. This indicates that, under the most optimistic estimation, we have verified that the gravitational sirens from a new information channel can effectively elucidate the current cosmological anomaly. Subsequently, we find that 1000 events can provide the matter density parameter $\Omega_{m0}=0.2872\pm0.0095$ with a little higher accuracy (3.3$\%$) than Planck analysis $\Omega_{m0}=0.308\pm0.012$ (3.9$\%$) \cite{11} (see the left panel of Fig. \ref{f5}). Using only Planck-2015 temperature angular power spectrum (Planck TT) and low-multipole polarization (lowP) data, the Planck Collaboration gives the constraint on the EoS of DE $\omega=-1.54^{+0.19}_{-0.39}$ at the $68\%$ CL \cite{11}. We find that about 800 events can provide the prediction of $\omega$ comparable to the Planck result and 1000 events give the constraint $\omega=-1.15^{+0.39}_{-0.31}$ at the $68\%$ CL (see the right panel of Fig. \ref{f5}). Furthermore, we conclude that about 1000 GW events can give the same prediction as Planck CMB result.

To demonstrate this conclusion better and exhibit the capabilities of gravitational sirens in constraining the cosmological models, we show the 1-dimensional and 2-dimensional marginalized posterior distributions for $\omega$CDM and CPL models in Figs. \ref{f6}-\ref{f7} using the 1000 simulated events and Planck CMB data. In Fig. \ref{f6}, one can easily find that the constrained $H_0$ and $\omega$ from 1000 GW events are well comparable to the Planck analysis using Planck TT + lowP. The former gives a larger EoS of DE and a smaller expansion rate of the universe than the latter. Very interestingly, unlike Planck TT + lowP, $H_0$ is very weakly anti-correlated with $\omega$ using GW data. The same situation also occurs in the case of CPL model (see Fig. \ref{f7}). One can also find that the GW data gives a higher cosmic expansion rate than the CPL case utilizing Planck TT + lowP + BAO (baryonic acoustic oscillations). Meanwhile, we report the $95\%$ upper limit of $\omega_a<1.28$ from GW data, which is much larger than the prediction $\omega_a<-0.045$ from Planck TT + lowP + BAO \cite{11}. It is worth noting that the GW data also cannot provide the positively-correlated properties between $H_0$ and $\omega_a$ like Planck TT + lowP + BAO.

\section{Summary and discussions}
Until 2015, due to the fact there is no realistic GW events, theorists just use the simulated GW data based on future detectors such as Einstein Telescope \cite{15,16,17,18,19,20,21,22,23} and Laser Interferometric Space Antenna to explore the evolution of the universe \cite{38,39,40,41,42}. The detections of GW has opened a new era of multi-messenger astronomy. Recently, with the third GW event reported by the LIGO Collaboration, it is very timely and necessary to start directly from current data to make forecasted constraints on the cosmological parameters.

Under the optimistic assumption described above, we use the Monte Carlo simulation technique to generate the GW data. Taking advantage of MCMC method, we find that about 1000 events can give the predictions of the Hubble constant and present matter density parameter with the accuracy comparable to the Planck-2015 results. About 800 events can provide the constraint on the EoS of DE with the nearly same precision as the Planck-2015 analysis using Planck TT + lowP. We also find that the 1000 simulated GW events can alleviate effectively the current $H_0$ tension from 3.4$\sigma$ to 1.97$\sigma$ CL.

It should be stressed again that, before the simulation, we have assumed boldly the redshifts of 4 GW events can be determined accurately and they are exactly the best-fit redshift points. As mentioned above, a possible realization is that these 4 events have accompanying EM signals and their redshifts can be measured with ultrahigh precision, which current detectors cannot identify.

From Fig. \ref{f5}, one can observe that when simulating 1000 events, the present matter density parameter and EoS of DE are consistent with the Planck prediction at the 68$\%$ CL. However, if we simulate more GW events, they may tend to be lower and higher than the Planck result, respectively. With more forthcoming GW data, this issue should be discussed further.

From Figs. \ref{f6}-\ref{f7}, we find that the $\Lambda$CDM model is still favored at the $68\%$ CL in the GW information channel. Very interestingly, based on the simulated GW data, the expansion rate of the universe is insensitive to the EoS of DE in the $\omega$CDM model and two free parameters in the CPL one. This issue is also worth being investigated in the future.

With the coming of GW astronomy, we expect the combination of two information channels, GW and EM, can provide more abundant information for us about the evolution and structure formation of the universe.

\section*{Acknowledgements}
D. Wang thanks W. Kaiser, K. Murata, B. Ratra, F. Canfora, Y. Sun and Y. Yan for insightful communications. X. Meng thanks S. Odintsov and S. Ray for helpful discussions. This work acknowledges partial support from the National Natural Science Foundation of China.


\begin{thebibliography}{99}
\bibitem{1}
A. G. Riess {\it et al.} [Supernova Search Team], Astron. J. {\bf 116}, 1009 (1998).

\bibitem{2}
S. Perlmutter {\it et al.} [Supernova Cosmology Project], Phys. Rev. Lett. {\bf 83}, 670 (1999).

\bibitem{3}
C. L. Bennett {\it et al.} [WMAP Collaboration], Astrophys. J. Suppl. Ser. {\bf 208}, 20 (2013).

\bibitem{4}
P. Ade {\it et al.} [Planck Collaboration], Astron. Astrophys. {\bf 571}, A16 (2014).

\bibitem{5}
D. H. Weinberg {\it et al.}, Phys. Rep. {\bf 530}, 87 (2013).

\bibitem{6}
M. Kilbinger {\it et al.} Astron. Astrophys. {\bf 497}, 677 (2009).

\bibitem{7}
P. Ade {\it et al.} [Planck Collaboration], Astron. Astrophys. {\bf 594}, A13 (2016).

\bibitem{8}
S. Weinberg, Rev. Mod. Phy. {\bf 61}, 1 (1989).

\bibitem{9}
S. Tulin and H. Yu, arXiv:1705.02358 [hep-ph].

\bibitem{10}
A. G. Riess {\it et al.}, Astrophys. J. {\bf 826}, 56 (2016).

\bibitem{11}
N. Aghanim et al. [Planck Collaboration], Astron. Astrophys. {\bf 596}, A107 (2016).

\bibitem{12}
R. Battye, T. Charnock and A. Moss, Phys. Rev. D {\bf 91}, 103508 (2015).

\bibitem{13}
E. Macaulay, I. K. Wehus and H. K. Eriksen, Phys. Rev. Lett. {\bf 111}, 161301 (2013).

\bibitem{14}
  B.~F.~Schutz,
  Nature {\bf 323}, 310 (1986).

\bibitem{15}
  D.~E.~Holz and S.~A.~Hughes,
  Astrophys.\ J.\  {\bf 629}, 15 (2005).



\bibitem{16}
  N.~Dalal, D.~E.~Holz, S.~A.~Hughes and B.~Jain,
  Phys.\ Rev.\ D {\bf 74}, 063006 (2006).

\bibitem{17}
  C.~L.~MacLeod and C.~J.~Hogan,
  Phys.\ Rev.\ D {\bf 77}, 043512 (2008).

\bibitem{18}
  C.~Cutler and D.~E.~Holz,
  Phys.\ Rev.\ D {\bf 80}, 104009 (2009).

\bibitem{19}
  B.~S.~Sathyaprakash, B.~F.~Schutz and C.~Van Den Broeck,
  Class.\ Quant.\ Grav.\  {\bf 27}, 215006 (2010).

\bibitem{20}
  S.~Nissanke, D.~E.~Holz, S.~A.~Hughes, N.~Dalal and J.~L.~Sievers,
  Astrophys.\ J.\  {\bf 725}, 496 (2010).

\bibitem{21}
  W.~Zhao, C.~Van Den Broeck, D.~Baskaran and T.~G.~F.~Li,
  Phys.\ Rev.\ D {\bf 83}, 023005 (2011).

\bibitem{22}
  S.~R.~Taylor, J.~R.~Gair and I.~Mandel,
  Phys.\ Rev.\ D {\bf 85}, 023535 (2012).

\bibitem{23}
  W.~Del Pozzo,
  Phys.\ Rev.\ D {\bf 86}, 043011 (2012).

\bibitem{24}
D. Markovic, Phys. Rev. D {\bf 48}, 4738 (1993).

\bibitem{25}
  C.~Messenger and J.~Read,
  Phys.\ Rev.\ Lett.\  {\bf 108}, 091101 (2012).

\bibitem{26}
  B.~P.~Abbott {\it et al.} [LIGO Scientific and Virgo Collaborations],
  Phys.\ Rev.\ Lett.\  {\bf 116}, 061102 (2016).

\bibitem{27}
  B.~P.~Abbott {\it et al.} [LIGO Scientific and Virgo Collaborations],
  Phys.\ Rev.\ Lett.\  {\bf 116}, 241103 (2016).

\bibitem{28}
  B.~P.~Abbott {\it et al.} [LIGO Scientific and Virgo Collaborations],
  Phys.\ Rev.\ Lett.\  {\bf 118}, 221101 (2017).

\bibitem{29}
  B.~P.~Abbott {\it et al.} [LIGO Scientific and Virgo Collaborations],
  Phys.\ Rev.\ D {\bf 93}, 122003 (2016).

\bibitem{30}
  B.~P.~Abbott {\it et al.} [LIGO Scientific and Virgo Collaborations],
  Phys.\ Rev.\ X {\bf 6}, 041015 (2016).

\bibitem{31}
  B.~P.~Abbott {\it et al.} [LIGO Scientific and Virgo Collaborations],
  Astrophys.\ J.\  {\bf 833}, L1 (2016).

\bibitem{32}
M. Chevallier, D. Polarski, Int. J. Mod. Phys. D {\bf 10}, 213 (2013).

\bibitem{33}
E. V. Linder, Phys. Rev. Lett. {\bf 90,} 091301 (2003).

\bibitem{34}
  V.~Connaughton {\it et al.},
  Astrophys.\ J.\  {\bf 826}, L6 (2016).

\bibitem{35}
  M.~Ackermann {\it et al.} [Fermi-LAT Collaboration],
  Astrophys.\ J.\  {\bf 823}, L2 (2016).

\bibitem{36}
  [Fermi-GBM and Fermi-LAT Collaborations],
  arXiv:1706.00199 [astro-ph.HE].

\bibitem{37}
A. Lewis and S. Bridle, Phys. Rev. D {\bf 66}, 103511 (2002).

\bibitem{38}
  A.~Petiteau, S.~Babak and A.~Sesana,
  Astrophys.\ J.\  {\bf 732}, 82 (2011).

\bibitem{39}
  N.~Tamanini {\it et al.},
  JCAP {\bf 1604}, 002 (2016).

\bibitem{40}
  K.~Kyutoku and N.~Seto,
  Phys.\ Rev.\ D {\bf 95}, 083525 (2017).

\bibitem{41}
  W.~Del Pozzo, A.~Sesana and A.~Klein,
  arXiv:1703.01300 [astro-ph.CO].

\bibitem{42}
  H.~Audley {\it et al.},
  arXiv:1702.00786 [astro-ph.IM].

























































































































\end{thebibliography}
\end{document}